\documentstyle[12pt,psfig]{article}
\makeatletter
\def\@cite#1#2{{[{#1}]\if@tempswa\typeout
{IJCGA warning: optional citation argument
ignored: `#2'} \fi}}
\newcount\@tempcntc
\def\@citex[#1]#2{\if@filesw\immediate\write\@auxout{\string\citation{#2}}\fi
  \@tempcnta\z@\@tempcntb\m@ne\def\@citea{}\@cite{\@for\@citeb:=#2\do
    {\@ifundefined
       {b@\@citeb}{\@citeo\@tempcntb\m@ne\@citea\def\@citea{,}{\bf ?}\@warning
       {Citation `\@citeb' on page \thepage \space undefined}}%
    {\setbox\z@\hbox{\global\@tempcntc0\csname b@\@citeb\endcsname\relax}%
     \ifnum\@tempcntc=\z@ \@citeo\@tempcntb\m@ne
       \@citea\def\@citea{,}\hbox{\csname b@\@citeb\endcsname}%
     \else
      \advance\@tempcntb\@ne
      \ifnum\@tempcntb=\@tempcntc
      \else\advance\@tempcntb\m@ne\@citeo
      \@tempcnta\@tempcntc\@tempcntb\@tempcntc\fi\fi}}\@citeo}{#1}}
\def\@citeo{\ifnum\@tempcnta>\@tempcntb\else\@citea\def\@citea{,}%
  \ifnum\@tempcnta=\@tempcntb\the\@tempcnta\else
   {\advance\@tempcnta\@ne\ifnum\@tempcnta=\@tempcntb \else \def\@citea{--}\fi
    \advance\@tempcnta\m@ne\the\@tempcnta\@citea\the\@tempcntb}\fi\fi}
\makeatother
\newenvironment{Eqnarray}%
     {\arraycolsep 0.14em\begin{eqnarray}}{\end{eqnarray}}

\def\simlt{\stackrel{<}{{}_\sim}}
\def\simgt{\stackrel{>}{{}_\sim}}
\def\be{\begin{equation}}
\def\ee{\end{equation}}
\def\bear{\be\begin{array}}
\def\eear{\end{array}\ee}
\def\bea{\begin{Eqnarray}}
\def\eea{\end{Eqnarray}}

\def\lsim{\mathrel{\raise.3ex\hbox{$<$\kern-.75em\lower1ex\hbox{$\sim$}}}}
\def\gsim{\mathrel{\raise.3ex\hbox{$>$\kern-.75em\lower1ex\hbox{$\sim$}}}}
\def\ifmath#1{\relax\ifmmode #1\else $#1$\fi}
\def\ls#1{\ifmath{_{\lower1.5pt\hbox{$\scriptstyle #1$}}}}

\def\beq{\begin{equation}}
\def\eeq{\end{equation}}
\def\beqa{\begin{Eqnarray}}
\def\eeqa{\end{Eqnarray}}

\def\baselinestretch{1}
\begin{document}
%%%%%%%%%%%%%%%%%%%%%%%%%%%%%%%%%%%%%%%%%%%%%%%%%%%%%%%%%%%%%
\def\IJMPA #1 #2 #3 {{\sl Int.~J.~Mod.~Phys.}~{\bf A#1}\ (19#2) #3$\,$}
\def\MPLA #1 #2 #3 {{\sl Mod.~Phys.~Lett.}~{\bf A#1}\ (19#2) #3$\,$}
\def\NPB #1 #2 #3 {{\sl Nucl.~Phys.}~{\bf B#1}\ (19#2) #3$\,$}
\def\PLB #1 #2 #3 {{\sl Phys.~Lett.}~{\bf B#1}\ (19#2) #3$\,$}
\def\PR #1 #2 #3 {{\sl Phys.~Rep.}~{\bf#1}\ (19#2) #3$\,$}
\def\JHEP #1 #2 #3 {{\sl JHEP}~{\bf #1}~(19#2)~#3$\,$}
\def\PRD #1 #2 #3 {{\sl Phys.~Rev.}~{\bf D#1}\ (19#2) #3$\,$}
\def\PTP #1 #2 #3 {{\sl Prog.~Theor.~Phys.}~{\bf #1}\ (19#2) #3$\,$}
\def\PRL #1 #2 #3 {{\sl Phys.~Rev.~Lett.}~{\bf#1}\ (19#2) #3$\,$}
\def\RMP #1 #2 #3 {{\sl Rev.~Mod.~Phys.}~{\bf#1}\ (19#2) #3$\,$}
\def\ZPC #1 #2 #3 {{\sl Z.~Phys.}~{\bf C#1}\ (19#2) #3$\,$}
\def\PPNP#1 #2 #3 {{\sl Prog. Part. Nucl. Phys. }{\bf #1} (#2) #3$\,$}
%%%%%%%%%%%%%%%%%%%%%%%%%%%%%%%%%%%%%%%%%%%%%%%%%%%%%%%%%

%%%%%%%%%%%%%%%%%%%%%%%%%%% subequations.sty %%%%%%%%%%%%%%%%%%%%%%%%
\catcode`@=11
\newtoks\@stequation
\def\subequations{\refstepcounter{equation}%
\edef\@savedequation{\the\c@equation}%
  \@stequation=\expandafter{\theequation}%   %only want \theequation
  \edef\@savedtheequation{\the\@stequation}% % expanded once
  \edef\oldtheequation{\theequation}%
  \setcounter{equation}{0}%
  \def\theequation{\oldtheequation\alph{equation}}}
\def\endsubequations{\setcounter{equation}{\@savedequation}%
  \@stequation=\expandafter{\@savedtheequation}%
  \edef\theequation{\the\@stequation}\global\@ignoretrue

\noindent}
\catcode`@=12
%%%%%%%%%%%%%%%%%%%%%%%%%%%%%%%%%%%%%%%%%%%%%%%%%%%%%%%%%%%%%%%%%%%%%
\begin{titlepage}

\title{{\bf  Naturalness of nearly degenerate neutrinos}}
\vskip2in
\author{ 
{\bf J.A. Casas$^{1,2}$\footnote{\baselineskip=16pt E-mail: {\tt
casas@mail.cern.ch}}}, 
{\bf J.R. Espinosa$^{2}$\footnote{\baselineskip=16pt E-mail: {\tt
espinosa@mail.cern.ch}}}, 
{\bf A. Ibarra$^{1}$\footnote{\baselineskip=16pt  E-mail: {\tt
alejandro@makoki.iem.csic.es}}} and 
{\bf I. Navarro$^{1}$\footnote{\baselineskip=16pt Email: {\tt
ignacio@makoki.iem.csic.es}}}\\ 
\hspace{3cm}\\
%\vskip.35in   
 $^{1}$~{\small Instituto de Estructura de la materia, CSIC}\\
 {\small Serrano 123, 28006 Madrid}
\hspace{0.3cm}\\
 $^{2}$~{\small Theory Division, CERN}\\
{\small CH-1211 Geneva 23, Switzerland}.
}
\date{}
\maketitle
\def\baselinestretch{1.15}
\begin{abstract}
\noindent
If neutrinos are to  play a relevant cosmological role, they must be
essentially degenerate. We study whether radiative corrections
can or cannot be responsible for the small mass splittings, in   
agreement  with all the available experimental data.
We perform an exhaustive exploration of the bimaximal mixing scenario,
finding that (i) the vacuum oscillations solution to the solar
neutrino problem is always excluded; (ii) if the mass matrix is
produced by a see-saw mechanism, there are large regions of the parameter
space
consistent with the large angle MSW solution, providing a natural origin
for the $\Delta m^2_{sol} \ll \Delta m^2_{atm}$ hierarchy; (iii) 
the bimaximal structure becomes then stable under radiative corrections. 
We also provide analytical
expressions for the mass splittings and mixing angles and present
a particularly simple see-saw ansatz consistent with all observations.
\end{abstract}

\thispagestyle{empty}
\leftline{}
\leftline{CERN-TH/99-103}
\leftline{April 1999}
\leftline{}

\vskip-20cm
\rightline{}
\rightline{IEM-FT-191/99}
\rightline{CERN-TH/99-103}
\rightline{IFT-UAM/CSIC-99-15}
\rightline{hep-ph/9904395}
\vskip3in

\end{titlepage}
%%%%%%%%%%%%%%%%%%%%%%%%%%%%%%%%%%%%%%%%%%%%%%%%%%%%%%%%%%%%%%%%%%%
\setcounter{footnote}{1} \setcounter{page}{1}
\newpage
%
% BODY
\baselineskip=20pt

\noindent

%{\bf 1.}~~

\section{Introduction}

Standard explanations of the observed atmospheric and
solar neutrino  anomalies~\cite{SK} require neutrino oscillations
between different species, which imply that neutrinos are massive,
with mass-squared differences of at most $10^{-2}\ {\rm eV}^2$.  On
the other hand, if neutrinos are to play an essential role  in the
large scale structure of the universe, the sum of their masses must be
a few eV, and therefore they must be almost degenerate.
This scenario has recently attracted much 
attention~\cite{GG}-\cite{degenerofilia}.  
In
this paper we will analyze under which circumstances the ``observed''
mass differences arise naturally (or not), as a radiative effect, in
agreement  with all the available experimental data.

\vspace{0.2cm} Let us briefly review the current relevant experimental
constraints on neutrino masses and mixing angles. Observations of
atmospheric neutrinos require $\nu_\mu-\nu_\tau$ oscillations driven
by a mass splitting and a mixing angle in the range \cite{range}
\bea 5\times10^{-4}\ {\rm eV}^2 <  &\Delta m^2_{at}& < 10^{-2}\ {\rm
eV}^2\ , \nonumber\\ \sin^22\theta_{at}&>&0.82\ .
\label{atm}
\eea
On the other hand, there are three main explanations of the
solar
neutrino flux deficits, requiring oscillations of electron neutrinos
into other species. The associated mass splittings and mixing
angles depend on the type of solution:

\noindent
\vspace{0.2cm} Small angle MSW (SAMSW) solution:
\bea 3\times10^{-6}\ {\rm eV}^2 <  &\Delta m^2_{sol}& < 10^{-5}\ {\rm
eV}^2, \nonumber\\ 4\times10^{-3} < &\sin^22\theta_{sol}& <
1.3\times10^{-2}.
\label{SAMSW}
\eea
\vspace{0.2cm} Large angle MSW (LAMSW) solution:
\bea 10^{-5}\ {\rm eV}^2 <  &\Delta m^2_{sol}& < 2\times 10^{-4}\ {\rm
eV}^2, \nonumber\\ 0.5 < &\sin^22\theta_{sol}& < 1.
\label{LAMSW}
\eea
\vspace{0.2cm} Vacuum oscillations (VO) solution:
\bea 5\times10^{-11}\ {\rm eV}^2 <  &\Delta m^2_{sol}& <
1.1\times10^{-10}\ {\rm eV}^2, \nonumber\\ \sin^22\theta_{sol} &>& 0.67\, .
\label{VO}
\eea
Let us remark the hierarchy of  mass splittings between the different
species of neutrinos, $\Delta m^2_{sol}\ll\Delta m^2_{at}$, which is
apparent from eqs.(\ref{atm}, \ref{SAMSW}--\ref{VO}). This hierarchy
should be reproduced by any natural explanation of those splittings.

As it has been shown in ref.~\cite{GG} the small angle MSW solution is
unplausible in a scenario of nearly degenerate neutrinos, so we are
left with the LAMSW and VO possibilities.

An important point concerns the upper limit on $\sin^22\theta_{sol}$
in the LAMSW solution. According to ref.\cite{Giunti}  the
absolute limit $\sin^22\theta_{sol}=1$ is  forbidden at 99.8\%, but
with  no indication about the tolerable upper limit. In this sense a
conservative upper bound $\sin^22\theta_{sol}<0.99$ can be adopted.
However, recent combined analysis of data,
including the day--night effect \cite{upper}, indicate that even the
$\sin^22\theta_{sol}=1$ possibility is allowed at 99\% for $2\times
10^{-5}\ {\rm eV}^2 <   \Delta m^2_{sol} < 1.7\times 10^{-4}\ {\rm
eV}^2$, although there may be problems to define the upper limit
on $\sin^22\theta_{sol}$ in a precise sense \cite{pc}.  
For the moment we will not adopt an upper
bound on $\sin^22\theta_{sol}$, but  later on we will show the effect
of such a bound on the results, which turns out to be dramatic.

Finally, the non-observation of neutrinoless double $\beta$-decay puts
an upper bound on the $ee$ element of the Majorana neutrino mass
matrix, namely ~\cite{Baudis}
\bea {\cal M}_{ee}<B=0.2\ {\rm eV}.
\label{B}
\eea

Concerning the cosmological relevance of neutrinos, as mentioned
above, we will assume $\sum m_{\nu_i} = O({\rm eV})$. In particular, we will
take $\sum m_{\nu_i} = 6$ eV as a typical possibility, although as we
will see, the results do not depend essentially on the particular
value. It should be mentioned here that Tritium $\beta$-decay
experiments indicate $m_{\nu_i} < 2.5$ eV for any mass eigenstate with
a significant $\nu_e$ component \cite{triti}.

Using standard notation, we define the effective mass term for the
three light (left-handed) neutrinos in the flavour basis as
\bea {\cal L}=-\frac{1}{2} \nu^T {\cal M_\nu}  \nu\;+\;{\rm h.c.}
\label{Mnu}
\eea
where ${\cal M_\nu}$ is the neutrino mass matrix. This is diagonalized
in the usual way
\bea {\cal M_\nu} = V^* D\, V^\dagger,
\label{V}
\eea
where $V$ is a unitary `CKM' matrix, relating the flavor
eigenstates to the mass eigenstates \bea \pmatrix{\nu_e \cr \nu_\mu\cr
\nu_\tau\cr}= \pmatrix{c_2c_3 &   c_2s_3 &   s_2e^{-i\delta}\cr
-c_1s_3-s_1s_2c_3e^{i\delta} &   c_1c_3-s_1s_2s_3e^{i\delta} &
s_1c_2\cr s_1s_3-c_1s_2c_3e^{i\delta} &   -s_1c_3-c_1s_2s_3e^{i\delta}
&   c_1c_2\cr}\, \pmatrix{\nu_1\cr \nu_2\cr \nu_3\cr}\,,
\label{CKM}
\eea
where $s_i$ and $c_i$ denote $\sin\theta_i$ and $\cos\theta_i$
respectively. $D$ may be written as \bea
D=\pmatrix{m_1e^{i\phi}&0&0\cr 0&m_2e^{i\phi'}&0\cr 0&0&m_3\cr}\, . \eea
It should be mentioned here that, for a given mass matrix, the
$\theta_i$ angles are not uniquely defined, unless one gives a
criterion to order the mass eigenvectors $\nu_i$ in eq.(\ref{CKM})
(e.g. $m_{\nu_1}^2<m_{\nu_2}^2<m_{\nu_3}^2$). Of course, the corresponding
$V$ matrices differ just in the ordering of the columns, and thus are
physically equivalent.

In this notation, constraint (\ref{B}) reads \bea {\cal M}_{ee} \equiv
\big\vert  m_{\nu_1}\,c_2^2c_3^2e^{i\phi} +
m_{\nu_2}\,c_2^2s_3^2e^{i\phi'} + m_{\nu_3}\, s_2^2\,e^{i2\delta}
\big\vert < B \,.
\label{doublebeta}
\eea As it has been put forward by Georgi and Glashow in ref.~\cite{GG}, a
scenario of nearly
degenerate neutrinos plausibly leads to $\theta_2\simeq 0$. In that
case, and for $m_\nu=2$ eV, eq.(\ref{doublebeta}) 
yields $\sin^2 2\theta_3\geq 0.99$,
which, as discussed above, might be in conflict with the
LAMSW solution to the solar neutrino anomaly. 
However,
according to some fits, $\theta_2$ could be as large as $27^0$ or even
larger \cite{range}. Therefore, although small values of $\theta_2$ are
clearly preferred, in some acceptable cases a non-negligible contribution of
$\sin^2 \theta_2$ in eq.(\ref{doublebeta}) is enough to relax the
above mentioned stringent bound on $\sin^2 2\theta_3$, and in fact we
will see some examples of this feature later on. Our criterion
throughout the paper will be to keep eq.(\ref{B}) [or the complete
eq.(\ref{doublebeta})] as the neutrinoless double $\beta$-decay
constraint, without demanding any extra condition on $\theta_3$,
$\theta_2$ \cite{vissani}. In any case, we will see that in all viable cases
$\sin^2 2\theta_3\geq 0.99$ and $\theta_2$ is very small.

\vspace{0.2cm} Let us now discuss the strategy we have followed to
analyze if the required mass splittings and mixing angles can or
cannot arise in a natural way through radiative corrections.

Following ref.~\cite{GG} (some of their arguments have been discussed
above),  the scenario of nearly degenerate neutrinos should be close
to a bimaximal mixing, which constrains the texture of the mass matrix
${\cal M}_\nu$ to be~\cite{GG,barger}  
\bea 
{\cal M}_{b} = m_\nu \,\pmatrix{
0 & {\displaystyle{1\over\sqrt2}} &  {\displaystyle{1\over\sqrt2}}\cr   {\displaystyle{1\over\sqrt2}} &
 {\displaystyle{1\over2}} &
- {\displaystyle{1\over2}}\cr  {\displaystyle{1\over\sqrt2}} &  - {\displaystyle{1\over2}} &
 {\displaystyle{1\over2}}\cr }\;,
\label{MGG}
\eea where $m_\nu$ is a general mass scale.  ${\cal M}_{b}$ can be
diagonalized by a $V$ matrix  \bea {V}_{b} =  \,\pmatrix{
 {\displaystyle{-1\over\sqrt2}} &   {\displaystyle{1\over\sqrt2}} &  0 \cr  {\displaystyle{1\over 2}}
&  {\displaystyle{1\over2}} &
 {\displaystyle{-1\over\sqrt2}} \cr  {\displaystyle{1\over2}} &  {\displaystyle{1\over2}} &
   {\displaystyle{1\over\sqrt2}} \cr }\;,
\label{VGG}
\eea leading to exactly degenerate neutrinos: $D=m_\nu\  {\rm diag}(-1, 1,
1)$ and $\theta_2=0$, $\sin^2 2\theta_3=\sin^2 2\theta_1=1$.
%The angles $\theta_1, \theta_3$ should be identified with the
%relevant ones for atmospheric and solar neutrino oscillations
%respectively. 
Let us remind that in this scenario only the LAMSW and
VO solutions to the solar anomaly are acceptable.

The nice aspect of ${\cal M}_{b}$ suggests that it could be generated
at some scale by interactions obeying appropriate continuous or
discrete symmetries. This is an interesting issue \cite{sym}, which we
will not address in this paper. On the other hand, in order to be
realistic, the effective matrix ${\cal M}_\nu$ at low energy should be
certainly close to ${\cal M}_{b}$, but it must be {\em slightly}
different in order to account for the mass splittings given in
eqs.(\ref{atm},\ref{LAMSW},\ref{VO}). The main goal of this paper is
to explore whether the appropriate splittings (and mixing angles) can
be generated or not through radiative corrections; more precisely,
through the running of the renormalization group equations (RGEs)
between the scale at which ${\cal M_\nu}$ is generated and low
energy. The output of this analysis can be of three types:

\begin{description}

\item[{\em i)}] All the mass splittings and mixing angles obtained
from the RG running are in agreement with all  experimental limits and
constraints.

\item[{\em ii)}] Some (or all) mass splittings are much larger than
the acceptable ranges.

\item[{\em iii)}] Some (or all) mass splittings are smaller than the
acceptable ranges, and the rest is within.

\end{description}

Case {\em (i)} is obviously fine. Case {\em (ii)} is
disastrous. The only way-out would be an extremely artificial
fine-tuning between  the initial form of ${\cal M}_\nu$ and the effect
of the RG running.  Hence we consider this possibility unacceptable.
Finally, case {\em (iii)} is not fine, in the sense that the RGEs fail
to explain the required modifications of ${\cal M}_\nu$. However, it
leaves the door open to the possibility that other (not specified)
effects could be responsible for them. In that case, the RGE would not
spoil a fine pre-existing structure. Consequently, we consider this
possibility as undecidable.
We will see along the paper different scenarios corresponding to the
three possibilities.

Concerning the mixing angles, it is worth stressing that, due to the
two degenerate eigenvalues of ${\cal M}_{b}$, ${V}_{b}$ is not
uniquely defined (${V}_{b}$ times any rotation on the plane of
degenerate eigenvalues is equally fine). Hence, once the ambiguity is
removed thanks to the small splittings coming from the RG running,
the mixing angles may be very different from the desired
ones. However, if those cases correspond to the previous {\em (iii)}
possibility, they are still of the ``undecidable'' type with respect
to the mixing angles, since the modifications of ${\cal M}$ (of
non-specified origin) needed to reproduce the correct mass
splittings will also change dramatically the mixing angles.

In section 2, we first examine the general case in which the neutrino masses
arise from an effective operator, remnant from new physics entering at a
scale $\Lambda$. In this framework, we take a low energy point of
view, assuming a bimaximal-mixing mass structure at the scale
$\Lambda$ as an initial condition. In this section we do not consider possible
perturbations of that initial condition coming from high-energy effects. 
We find this case to be of the
undecidable type [possibility {\em (iii)} above], except for the VO solution,
which is excluded.

In section 3 we go one step further and consider in detail a particularly well
motivated example of the previous case: the see-saw scenario. We include here
the high energy effects of the new degrees of freedom above the scale $\Lambda$
(identified now with the mass of the right-handed neutrinos). 
We find regions of parameter space
where the neutrino spectrum and mixing angles fall naturally in the 
pattern required to explain solar (LAMSW solution) and atmospheric neutrino
anomalies, which we find remarkable. We complement the numerical results
with compact analytical formulas which give a good description of them, and
allow to understand the pattern of mass splittings and mixing angles obtained.
We also present  plausible textures for the neutrino Yukawa couplings leading
to a good fit of the oscillation data. Finally we draw some conclusions.

\section{${\cal M}_\nu$ as an effective operator}

In this section we consider the simplest possibility that the Majorana
mass matrix for the left-handed neutrinos, ${\cal M}_\nu$, is
generated at some high energy scale, $\Lambda$, by some unspecified
mechanism (we allow $\Lambda$ to vary from $M_p$ to $M_Z$). Assuming that
the only light fields below $\Lambda$ are the
Standard Model (SM) ones, the lowest dimension operator producing a
mass term of this kind is uniquely given by \cite{eff}
\bea -\frac{1}{4}\kappa \nu^T   \nu H H \;+\;{\rm h.c.}
\label{kappa}
\eea
where $\kappa$ is a matricial coupling and $H$ is the ordinary
(neutral) Higgs. Obviously, ${\cal
M}_\nu=\frac{1}{2}\kappa \langle H\rangle^2$. The effective coupling
$\kappa$ runs with the scale below $\Lambda$, with a RGE given by
\cite{Babu}
\bea 16\pi^2 \frac{d \kappa}{dt}= \left[-3g_2^2+2\lambda+6Y_t^2+2 {\rm
Tr}{\bf Y_e^\dagger Y_e} \right]\kappa -\frac{1}{2}\left[\kappa{\bf
Y_e^\dagger Y_e} + ({\bf Y_e^\dagger Y_e})^T\kappa\right]\ ,
\label{rg1}
\eea where $t=\log \mu$, and $g_2,\lambda, Y_t, {\bf Y_e}$ are the
$SU(2)$ gauge coupling, the quartic Higgs coupling, the top Yukawa
coupling and the matrix of Yukawa couplings for the charged leptons,
respectively. Let us note that the RGE depends on $\lambda$, 
and thus on the value of the Higgs mass, $m_H$. We have taken $m_H=150$ 
GeV throughout the paper, but in any case the dependence is very small 
(it slightly affects the overall neutrino mass scale but not the relative
splittings).

In the scenario in which we are interested (almost degenerate
neutrinos), the simplest assumption about the form of $\kappa$ is that
the interactions responsible for it produce the bimaximal mixing
texture of eq.(\ref{MGG}). Hence
\bea {\cal M}_\nu(\Lambda)=\frac{1}{2}\kappa(\Lambda) \langle
H\rangle^2={\cal M}_{b}\, .
\label{kappaboundary}
\eea
Clearly, the last term of the RGE (\ref{rg1}) will slightly
perturb the initial form of  ${\cal M}_{\nu}$, so we expect at low
energy small mass splittings, and mixing angles different from the
bimaximal case.

In order to gain intuition on the final (low energy) form of ${\cal
M}_{\nu}$, and the corresponding mass splittings and mixing angles, it
is convenient to neglect for a while all the charged lepton Yukawa
couplings but $Y_\tau$. Then $\kappa$ maintains its form along the
running, except for the third row and column:
\bea {\cal M}_\nu(\mu_0)\propto\pmatrix{ 0 &  {\displaystyle{1\over\sqrt2}} &
 {\displaystyle{1\over\sqrt2}}(1+\epsilon)\cr  {\displaystyle{1\over\sqrt2}} &
{\displaystyle{1\over2}}
&
- {\displaystyle{1\over2}}(1+\epsilon)\cr  {\displaystyle{1\over\sqrt2}}(1+\epsilon) &
- {\displaystyle{1\over2}}(1+\epsilon) &   
{\displaystyle{1\over2}}(1+2 \epsilon)\cr }\,
\label{Mpert}
\eea
where $\mu_0$ is the low-energy scale, which can be identified with
$M_Z$ and $\epsilon$ has positive sign.  
Therefore, the mass eigenvalues are proportional to
$1, \ -1-{\epsilon}/{2}, \ 1+{3\epsilon}/{2}$, and the
corresponding splittings (adopting the convention 
$m_{\nu_1}^2<m_{\nu_2}^2<m_{\nu_3}^2$) are
\bea \Delta m^2_{12}\ =\ \frac{1}{2}\Delta m^2_{23}\ =\
\frac{1}{3}\Delta m^2_{13} \ \simeq\  m_\nu^2 {\epsilon}.
\label{Deltaskappa}
\eea
%
%%%%%%%%%%%%%%%%%%%%%%%%figure%%%%%%%%%%%%%%%%%%%%%%%%
\begin{figure}[t]  
%\psdraft
\centerline{
\psfig{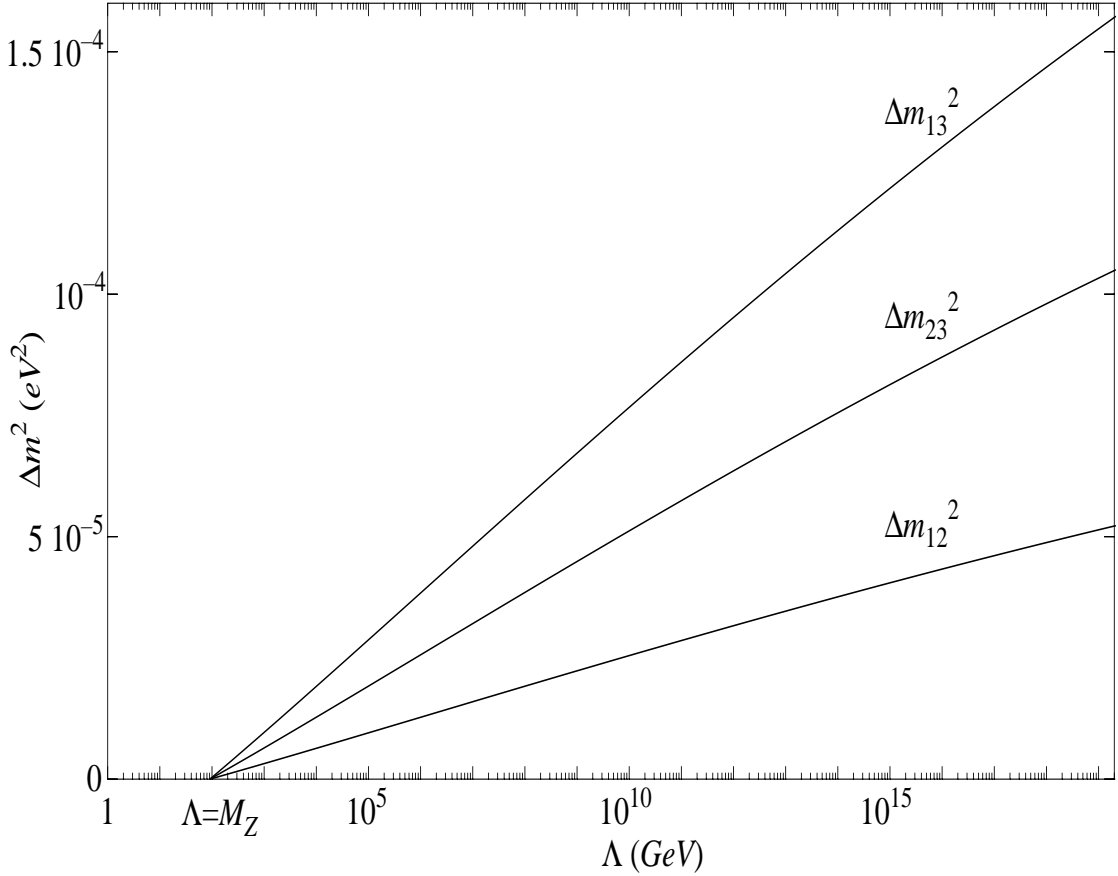}}
\caption
{\footnotesize Dependence of neutrino mass
splittings at
low energy ($\Delta m_{ij}^2$ in ${\mathrm eV}^2$) with the cut-off scale 
$\Lambda ({\rm GeV}$).
}
\end{figure}
%%%%%%%%%%%%%%%%%%%%%%%%figure%%%%%%%%%%%%%%%%%%%%%%%%

Clearly, these
mass
splittings are incompatible with the required
hierarchy $\Delta m^2_{sol}\ll\Delta m^2_{at}$, of eqs.(\ref{atm},
\ref{LAMSW}, \ref{VO}).  We will discuss the size of these splittings
shortly.  Equation (\ref{Mpert}) is also useful to get an approximate
form of the $V$ matrix responsible for the diagonalization of ${\cal
M}_{\nu}$
\bea V\simeq\pmatrix{  
{\displaystyle{1\over\sqrt3}} &
-{\displaystyle{1\over\sqrt2}} &   
 {\displaystyle{1\over\sqrt6}}\cr  
 {\displaystyle{\sqrt{2\over3}}} &
{\displaystyle{1\over2}} &  
- {\displaystyle{1\over2\sqrt3}}\cr  
0 &  
{\displaystyle{1\over2}} &  
{\displaystyle {\sqrt3\over2}}\cr
}\, ,
\label{Vkappa}
\eea
which leads to mixing angles
\bea \sin^2 2 \theta_1 = {9\over25},\;\;\; \sin^2 2 \theta_2 =
{5\over9},\;\;\; \sin^2 2 \theta_3 =  {24\over25}.
\label{mixingsskappa}
\eea
Clearly, these values are far away from the bimaximal mixing ones.
In consequence, they are not acceptable.

The previous failures of the scenario considered in this section in
order to reproduce the mass splittings and the mixing angles indicate
that we are in one of the two possibilities {\em (ii)} or {\em (iii)}
discussed in the Introduction.  To go further, we need a numerical
evaluation of the mass splittings, and thus 
of $\epsilon$. Solving the
RGE (\ref{rg1}) at lowest order, we simply find
\bea
\epsilon =\frac{Y_\tau^2}{32\pi^2}
\log\frac{\Lambda}{\mu_0}\, .
\label{epsilonkappa}
\eea
So, from eq.(\ref{Deltaskappa}) the mass splittings are typically of
order $10^{-5}$ eV$^2$. This is confirmed by the complete numerical
evaluation  of the RGE , which gives the mass splittings shown in
Fig.1.

The first conclusion is that for any $\Lambda$ (even very close to
$\mu_0$) the mass splittings are much larger than those required for
the VO solution to the solar neutrino problem, $\Delta m^2_{sol}\sim
10^{-10}$ eV$^2$. Therefore, the effect of the RGE for
this scenario is disastrous in the sense discussed in the Introduction
for the possibility {\em (ii)}. In consequence the VO solution to the
solar neutrino problem is excluded.

For the LAMSW solution to the solar neutrino problem, things are a bit
different. Clearly, the mass splittings obtained from the RGE analysis
are within or below the required range, eq.(\ref{LAMSW}), for almost 
any value of the
cut-off scale $\Lambda$. In addition, the mass splittings are clearly
below the atmospheric range, eq.(\ref{atm}). Therefore, we are in the
case {\em (ii)} discussed in the Introduction: The radiative
corrections fail to provide an  origin to the required mass splittings
and mixing angles, but they will not destroy a suitable initial
modification (from unspecified origin) of ${\cal M}_\nu$ at the
$\Lambda$ scale. It is anyway remarkable that the radiative corrections
place $\Delta m^2$ just in the right magnitude for the LAMSW solution.

In the next section we will discuss a natural source for the initial
modification of ${\cal M}_\nu$, which leads naturally to
completely satisfactory  (atmospheric plus LAMSW) scenarios.

\section{${\cal M}_\nu$ from the see-saw mechanism}

A natural way of obtaining small neutrino masses is the so-called
see-saw mechanism \cite{seesaw} in which the particle content of the
Standard Model is enlarged by one additional neutrino field (not
charged under the SM group) per generation, $\nu_{\alpha,R}$
($\alpha=e,\mu,\tau$). The Lagrangian reads \bea {\cal L}=
-\bar{\nu}_R m_D \nu + \frac{1}{2} \bar{\nu}_R {\cal M} \bar{\nu}_R^T
+ {\mathrm  h.c.}  \eea Here $m_D$ is a $3\times 3$ Dirac mass matrix
with magnitude determined by the electroweak breaking scale,
$m_D={\bf Y_\nu}\langle H\rangle$, where ${\bf Y_\nu}$ is the matrix
of neutrino Yukawa couplings and $\langle H\rangle=246/\sqrt{2}$ GeV; and
${\cal M}$ is a $3\times 3$ Majorana mass matrix which does not break the
SM gauge symmetry. Thus, the overall scale of ${\cal M}$, which we will
denote  by $M$, can be
naturally  many orders of magnitude higher than the electroweak scale.
In that case, the low-energy effective theory, after integrating out
the heavy $\nu_{\alpha,R}$ fields [whose masses are $O(M)$], is just
the SM with left-handed neutrino masses given by \bea {\cal M}_\nu=
m_D^T {\cal M}^{-1} m_D =  {\bf Y_\nu}^T {\cal M}^{-1} {\bf Y_\nu}
\langle H\rangle^2, \eea suppressed with respect to the typical
fermion masses by the  inverse power of the large scale $M$.

In this appealing framework, the degeneracy in neutrino masses can
come about as a result of some symmetry (at some high energy,
e.g. $M_p$) in the textures of  ${\cal M}$ and ${\bf Y_\nu}$ such that
${\cal M}_\nu(M_p)={\cal M}_{b}$.  Starting from this very symmetric
condition at $M_p$ we make contact with the low energy neutrino mass
matrix using the renormalization group. First we run  ${\cal M}$ and
${\bf Y_\nu}$ from $M_p$ down to $M$, where the right-handed neutrinos are
decoupled. Below that scale and all the way down to low energy we run
${\cal M}_\nu$ as an effective operator, in the same way as in the
previous section. We can also think of the first stage of running
between $M_p$ and $M$ as the high energy effects providing an starting
form of  ${\cal M_\nu}$  for the second stage of running, i.e. the only one
considered in the previous section (the role of $\Lambda$ is played
by $M$). This form, being slightly different 
from ${\cal M}_{b}$, may rescue the previous ``undecidable'' cases.

In pursuing this idea, it is natural to assume \cite{altfe} that the
structure leading to the relation ${\cal M}_\nu(M_p)={\cal M}_{b}$
occurs either in the Dirac matrix $m_D$ [while the Majorana matrix
${\cal M}$ is simply diagonal with equal eigenvalues (up to a
sign)\footnote{Actually, the only condition we are imposing is that the
eigenvalues are equal (up to a sign). A suitable transformation will
diagonalize ${\cal M}$ to the form we assume.}] or in the Majorana matrix [with
a simple diagonal Dirac matrix (again with eigenvalues equal up to a
sign)]. We
cannot expect a conspiracy between
both matrices (which come from totally different physics) leading to
${\cal M}_\nu(M_p)={\cal M}_{b}$.

\subsection{Textures of ${\cal M}$ and ${\bf Y_\nu}$ leading to bimaximal
mixing}

In the analysis of what structures for ${\cal M}$ or ${\bf Y_\nu}$
lead to ${\cal M}_\nu(M_p)={\cal M}_{b}$ for simplicity we only consider
real matrices (thus avoiding potential problems with CP violation).

If we assume that ${\bf Y_\nu}$ is proportional to the identity and
all the structure arises from the Majorana mass matrix, then we simply
need to impose ${\cal M}(M_p)\propto  {\cal M}_{b}$ (note that ${\cal
M}_{b}^{-1}\propto {\cal M}_{b}$ ).  The alternative case is less
trivial but equally simple. If ${\cal M}$ is of the form\footnote{If ${\cal M}$ is
taken proportional to the identity, the
Yukawa matrix ${\bf Y_\nu}$ must be chosen complex. Our choice of
${\cal M}$ avoids that complication and is physically equivalent.}
\bea
{\cal M}=M\pmatrix{-1 & 0 & 0\cr
0 & 1 & 0\cr
0& 0& 1}\, ,
\label{Mmaj}
\eea
then it is easy to see that the most general form
of ${\bf Y_\nu}$ satisfying ${\bf Y_\nu}^T {\cal M}^{-1} {\bf Y_\nu} \propto
{\cal M}_{b}$ is 
\bea {\bf Y_\nu}= Y_\nu B V_{b}^T \, .
\label{Ynu}
\eea 
Here $Y_\nu$ is the overall magnitude of ${\bf Y_\nu}$ and $B$ is a
combination of two `boosts' 
\bea 
B=\pmatrix{ \cosh a & 0 &  \sinh a
\cr 0 &  1 &  0 \cr  \sinh a   &     0    & \cosh a  \cr} \pmatrix{
\cosh b   &  \sinh b   &   0     \cr  \sinh b  &  \cosh b   & 0 \cr 0
   &   0  & 1 \cr} \, ,
\label{boosts}
\eea 
with two free parameters $a,b$. Actually,
one could also take  ${\bf Y_\nu}=  Y_\nu RB V_{b}^T$, where $R$ is a
rotation in the $(\mu,\tau)$ plane, but such rotations can be
absorbed into a change of the $(\nu_R)_\alpha$ basis,
$\nu_R\rightarrow R\nu_R$, with no modification in ${\cal M}$ 
and thus give physically equivalent
results.

Let us notice that the former case, where all the structure is in the ${\cal
M}$
matrix, is equivalent to the latter case (all the structure in ${\bf Y_\nu}$)
if we set $a=b=0$. To see this, note that a redefinition of the $\nu_R$
fields 
as $\nu_R\rightarrow V_b\nu_R$, would leave ${\cal M}\propto{\rm diag}(-1,1,1)$
and ${\bf Y_\nu}=Y_\nu V_{b}^T$. In consequence, it is enough to
study 
the case where all the structure is in ${\bf Y_\nu}$, given by eq.(\ref{Ynu}).

\subsection{Running ${\cal M}_\nu$ from $M_p$ to low energy}

From $M_p$ to $M$ the evolution of the relevant matrices is
governed by the following renormalization group equations~\cite{RGE}: 
\bea
\label{rg2}
\frac{d {\bf Y_\nu}}{dt}= -\frac{1}{16\pi^2} {\bf Y_\nu}\left[\left(
\frac{9}{4}g_2^2+\frac{3}{4}g_1^2-{\mathrm T}
\right){\bf I_3}-\frac{3}{2}\left( {\bf
Y_\nu}^\dagger  {\bf Y_\nu}-{\bf Y_e^\dagger Y_e}\right) \right], \eea
\bea
\label{rg3}
\frac{d {\bf Y_e}}{dt}= -\frac{1}{16\pi^2} {\bf Y_e} \left[\left(
\frac{9}{4}g_2^2+\frac{15}{4}g_1^2-{\mathrm T}
\right){\bf I_3}+\frac{3}{2}\left(
{\bf
Y_\nu}^\dagger {\bf Y_\nu}- {\bf Y_e^\dagger Y_e}\right) \right], \eea 
where
\bea
{\mathrm T}={\mathrm Tr}(3{\bf Y_U^\dagger Y_U}
+3 {\bf Y_D^\dagger Y_D}+{\bf Y_\nu^\dagger Y_\nu}+{\bf Y_e^\dagger Y_e}),
\eea
and
\bea
\label{rg4}
\frac{d {\cal M}}{dt}=\frac{1}{16\pi^2}\left[{\cal M} ({\bf Y_\nu}
{\bf Y_\nu}^\dagger)^T+{\bf Y_\nu} {\bf Y_\nu}^\dagger {\cal M}\right]
\eea 
(not yet given in the literature). Here $g_2$ and $g_1$ are the
$SU(2)_L$ and $U(1)_Y$ gauge coupling constants, and ${\bf Y_{U,D,e}}$ are
the Yukawa matrices for up quarks, down quarks and charged leptons.

At $M$, $\nu_R$ decouple, and ${\bf Y_e}$ must be diagonalized to
redefine the flavour basis of leptons [note that the last term in
(\ref{rg3}) produces non-diagonal contributions to ${\bf Y_e}$] affecting the
form of the ${\bf Y_\nu}$ matrix. 
Then the effective mass matrix for
the light neutrinos is ${\cal M}_\nu\simeq {\bf Y_\nu}^T {\cal M}^{-1} {\bf
Y_\nu} \langle H\rangle^2$.

From $M$ to $M_Z$, the effective mass matrix ${\cal M}_\nu$ is
run down in energy exactly as described in section 2.

The renormalization group equations are
integrated with the following boundary conditions:
${\cal M}$ and ${\bf Y_\nu}$ are chosen at $M_p$ so as to satisfy \bea
{\cal M}_\nu(M_p)={\cal M}_{b}, \eea with the overall magnitude of
${\bf Y_\nu}$ fixed, for a given value of the Majorana mass $M$, by the
requirement $m_{\nu}\sim O({\rm eV})$.  The boundary conditions for
the other Yukawa couplings are also fixed at the low energy  side to give
the observed fermion masses.  The free parameters are therefore $M, a$ and $b$.

\subsection{Limits on the parameter space}

We discuss here the limits on the parameter space of our study, which is 
expanded by $M, a$ and $b$.

The parameters $a$ and $b$, which define the texture of the ${\bf Y_\nu}$
matrix through eqs.(\ref{Ynu}, \ref{boosts}), can be in principle any real
numbers. However, it is apparent from eqs.(\ref{Ynu}, \ref{boosts}) that
if
$a$ or $b$ are large, the entries of ${\bf Y_\nu}$ are extremely
fine-tuned. Notice that the relative difference between $\cosh a$ and
$\sinh a$ factors is $\sim 2e^{-2|a|}$ (and the analogue for $\cosh b$ and
$\sinh b$). Therefore, the ${\bf Y_\nu}$ matrix is fine-tuned as $\sim
2e^{-2(|a|+|b|)}$ parts in one. In particular, if $|a|$ or $|b|$ are
$>1.5$
the matrix elements are fine-tuned at least in a 10\%. In consequence, we
will demand \bea |a|, |b| \leq 1.5 \, .\label{ablimits} \eea Let us remark
that
the previous limits are based on a criterion of naturality for the ${\bf
Y_\nu}$ matrix. However, let us mention that if we relax these
limits, the final results (to be presented in the next subsection) are
basically unchanged, since the allowed regions for $a,b$ remain in the
``natural'' region (\ref{ablimits}) in most of the cases.

Concerning the remaining parameter, $M$, there is an upper bound on it coming
from the fact that for large values of $M$, the neutrino Yukawa couplings,
${\bf Y_\nu}$, develop Landau poles below $M_p$, spoiling the
perturbativity of the theory \cite{cdiq}. This occurs for $M(M_p)\sim
4.3\times10^{13}$
GeV. Actually, there is an additional effect, namely the closing
of the allowed Higgs window, which occurs approximately for the same value of
$M(M_p)$. Consequently, this sets the upper bound on $M$. It
is
interesting to note that this effect also restricts the values of $a, b$:
for a given $M$, the larger $a, b$, the larger the entries of ${\bf
Y_\nu}$, and thus the lower the scale at which the Landau pole appears. In
general, this restriction on $a, b$ is less severe than
eq.(\ref{ablimits}).

Regarding the lower bound on $M$, there is no physical criterion for it,
as
the actual origin of the right-handed Majorana matrix is unknown. Since
$M$
can be written in terms of $m_\nu$ and ${\bf Y_\nu}$ (roughly speaking $M
\simeq Y_\nu^2 \langle H\rangle^2/m_\nu$), we can adopt the sensible
criterion that $Y_\nu$ is at least as large as the smallest Yukawa
coupling
so far known, i.e. the electron one. This precisely corresponds to $M\sim  
100$ GeV, below which is unplausible to descend. In consequence, our
limits
for $M$ are 
\bea 10^2 {\rm GeV}\ \simlt M\ \simlt 4.3\times 10^{13} {\rm
GeV}.
\label{Mlimits} \eea 
On the other hand, as we will see in the next subsection, there are
no physically viable scenarios unless \bea
 M\ \simgt 10^{8} {\rm GeV} ,
\label{Mlimit}
\eea
which sets an operating lower limit on $M$.

%%%%%%%%%%%%%%%%%%%%%%%%figure%%%%%%%%%%%%%%%%%%%%%%%%
\begin{figure}[t]
%\psdraft
\centerline{\hbox{  
\psfig{figure=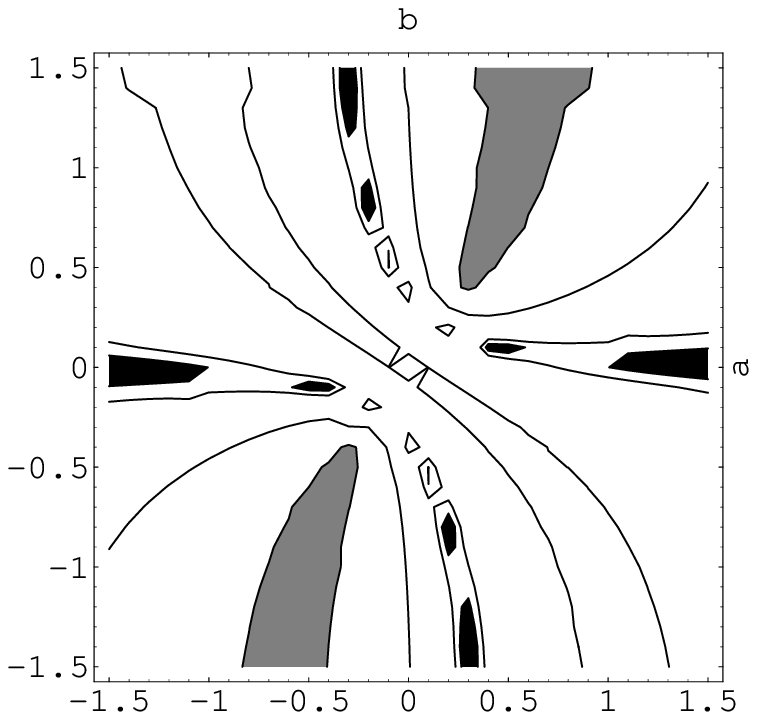,height=10cm,width=10cm,bbllx=1.cm,%
bblly=0.cm,bburx=12.cm,bbury=9.cm}
\psfig{figure=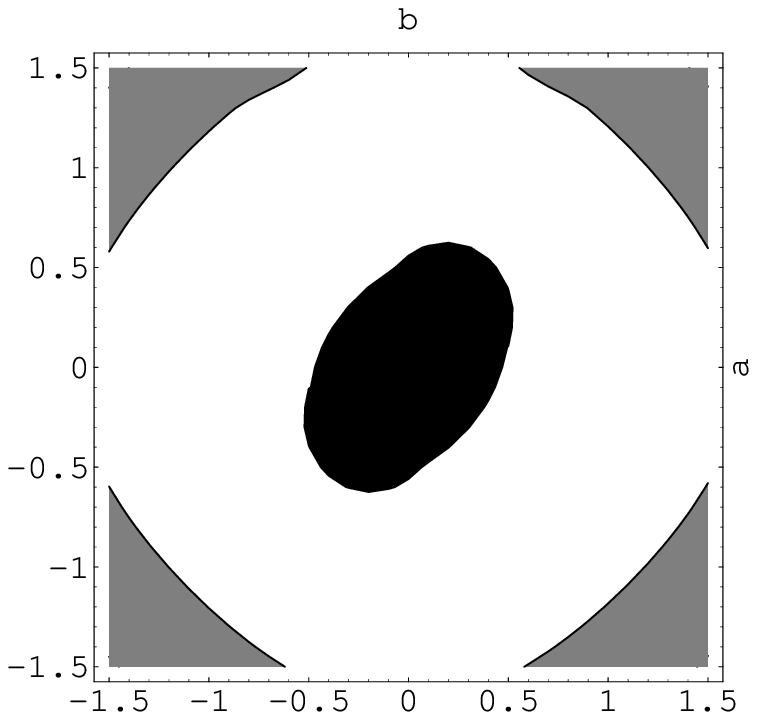,height=10cm,width=10cm,bbllx=4.cm,%
bblly=0.cm,bburx=15.cm,bbury=9.cm}}}
\caption{\footnotesize Left plot:
contours of $\Delta m_{12}^2/{\mathrm eV}^2$ in the $(b,a)$
plane from less than $10^{-5}$ (black area), through $2\times 10^{-5}$ and
$10^{-4}$ (lines) to more than $2\times 10^{-4}$ (grey).
Right plot: same for $\Delta m_{23}^2/{\mathrm eV}^2$, from
$5\times 10^{-4}$ (black) to $10^{-2}$ (grey). The Majorana mass is
$8\times 10^9$ GeV.}
\end{figure}
%%%%%%%%%%%%%%%%%%%%%%%%figure%%%%%%%%%%%%%%%%%%%%%%%%

\subsection{Results}

Figures 2 to 6 present our results for the mass splittings and mixing
angles at low energy, after numerical integration of the RGEs from $M_p$
to
low energy as described in subsection 3.2 (we follow the convention 
$m_{\nu_1}^2<m_{\nu_2}^2<m_{\nu_3}^2$ in all the figures).
We have chosen $M=8\times 10^9$
GeV and $m_\nu\simeq 2$ eV as a typical example; 
the dependence of the results with $M$ and $m_\nu$ is
discussed later on.

%%%%%%%%%%%%%%%%%%%%%%%%figure%%%%%%%%%%%%%%%%%%%%%%%%
\begin{figure}[t]
%%\psdraft
\centerline{\hbox{
\psfig{figure=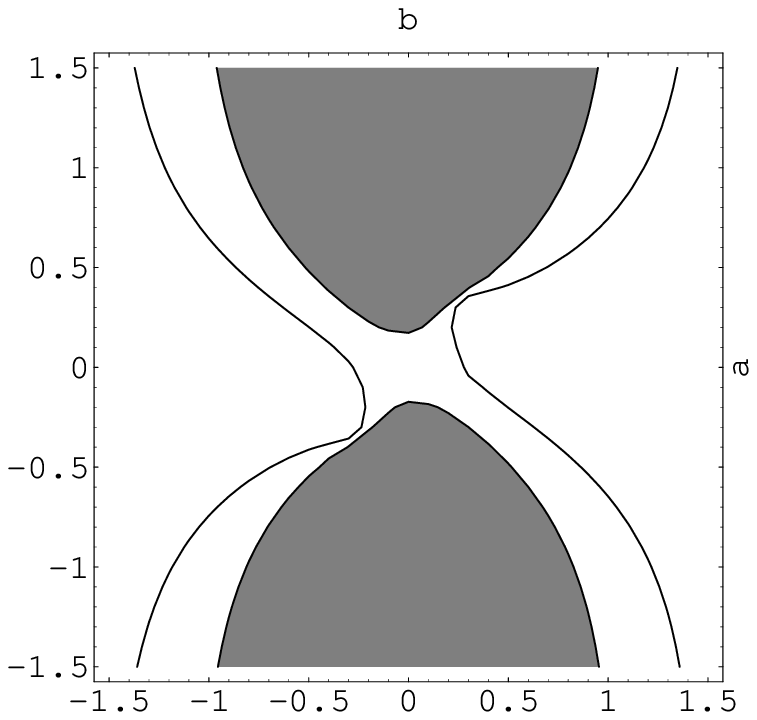,height=10cm,width=10cm,bbllx=1.cm,%
bblly=0.cm,bburx=12.cm,bbury=9.cm}
\psfig{figure=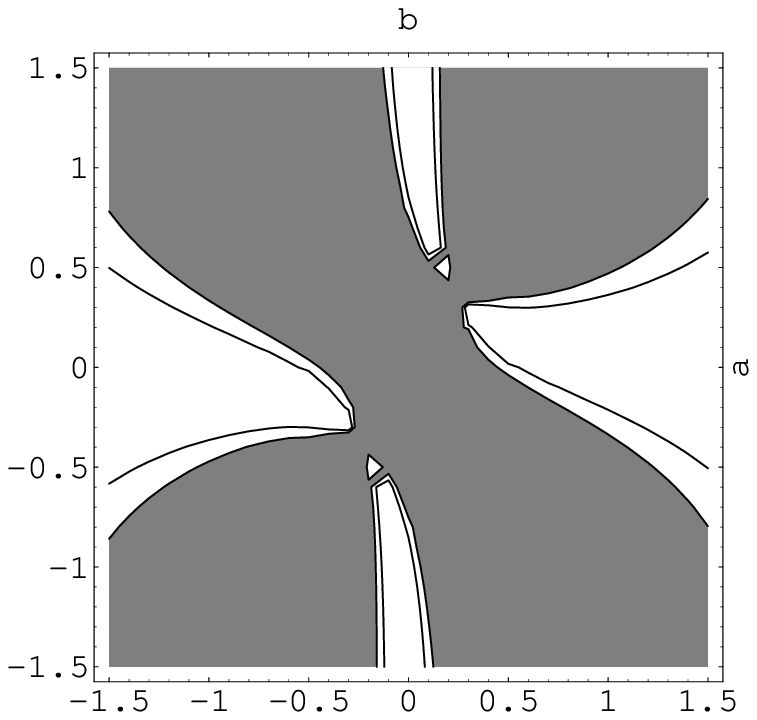,height=10cm,width=10cm,bbllx=4.cm,%
bblly=0.cm,bburx=15.cm,bbury=9.cm}}}
\caption{\footnotesize Left plot: Contours of
$\sin^22\theta_{2}$ in the $(b,a)$ plane. The grey area marks the
$\sin^22\theta_{2}>0.64$ region. The line singled-out corresponds to
$\sin^22\theta_{2}=0.36$. Right
plot: Contours of
$\sin^22\theta_{1}$ in the $(b,a)$ plane. In the grey area
$\sin^22\theta_{1}$ is smaller than 0.82, and the line corresponds to
$\sin^22\theta_{1}=0.9$. The Majorana mass is
$8\times 10^9$ GeV.}
\end{figure}
%%%%%%%%%%%%%%%%%%%%%%%%figure%%%%%%%%%%%%%%%%%%%%%%%%

Figure 2, left plot, shows contour lines of constant $\Delta m_{12}^2$ (the
squared
mass difference between the lightest neutrinos) in the plane $(b,a)$. The
black (grey) region is excluded because there $\Delta m_{12}^2<10^{-5}\,
{\mathrm eV}^2$ ($\Delta m_{12}^2>2\times 10^{-4}\, {\mathrm eV}^2$),
which is too
small
(large)
to account for the oscillations of solar neutrinos (LAMSW solution). The
white area is thus the allowed region. The lines in it correspond to
$\Delta m_{12}^2=10^{-4}\, {\mathrm eV}^2$ and $2\times 10^{-5}\,
{\mathrm eV}^2$.

%%%%%%%%%%%%%%%%%%%%%%%%figure%%%%%%%%%%%%%%%%%%%%%%%%
\begin{figure}[t]
%%\psdraft
\centerline{\hbox{
\psfig{figure=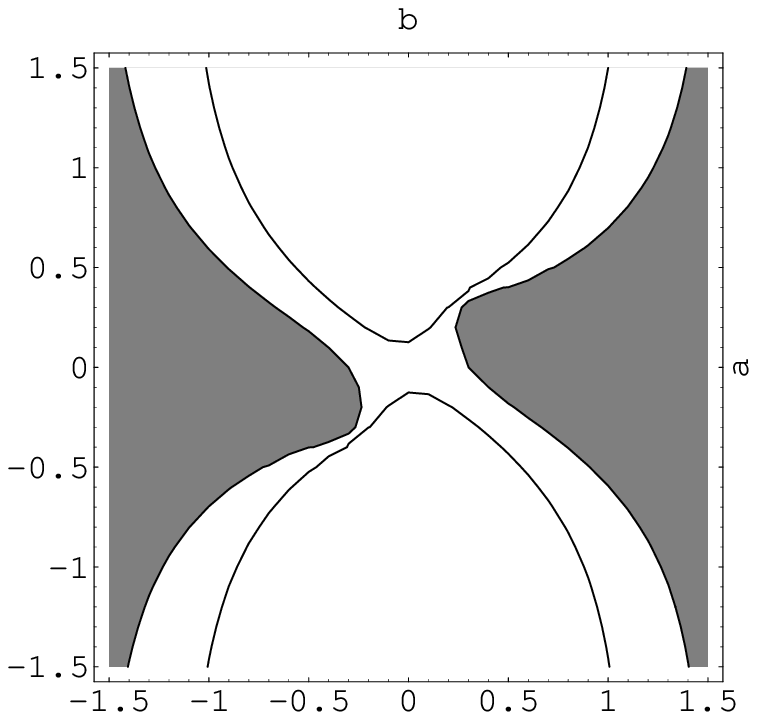,height=10cm,width=10cm,bbllx=1.cm,%
bblly=0.cm,bburx=12.cm,bbury=9.cm}
\psfig{figure=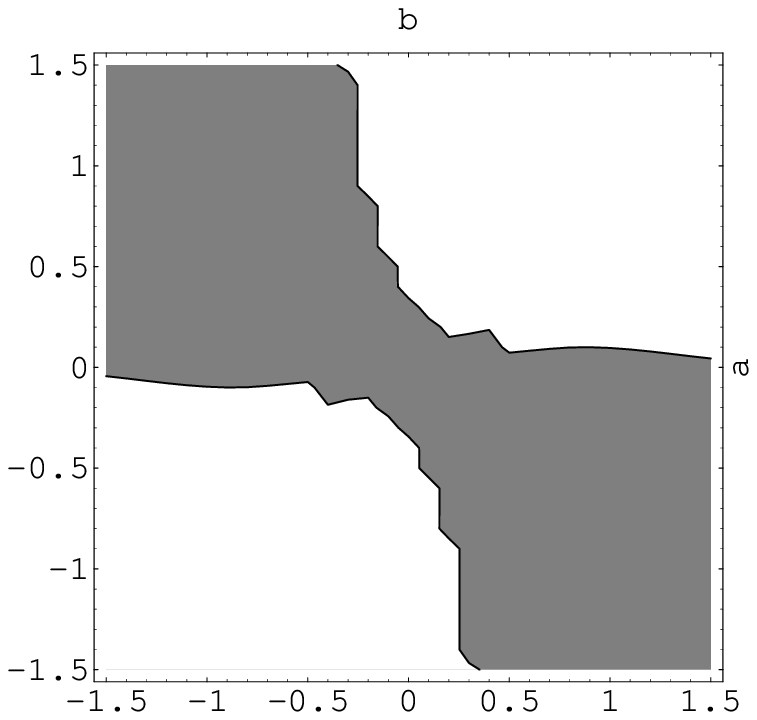,height=10cm,width=10cm,bbllx=4.cm,%
bblly=0.cm,bburx=15.cm,bbury=9.cm}}}
\caption{\footnotesize Left plot: Same as figure 3 for
$\sin^22\theta_{3}$. The grey area corresponds to values above 0.99.
The curve gives $\sin^22\theta_{3}=0.95$. Right plot: The grey area 
corresponds to $\cos 2\theta_{3}<0$
}
\end{figure}
%%%%%%%%%%%%%%%%%%%%%%%%figure%%%%%%%%%%%%%%%%%%%%%%%%

%%%%%%%%%%%%%%%%%%%%%%%%figure%%%%%%%%%%%%%%%%%%%%%%%%
\begin{figure}[t]
%%\psdraft
\centerline{
\psfig{figure=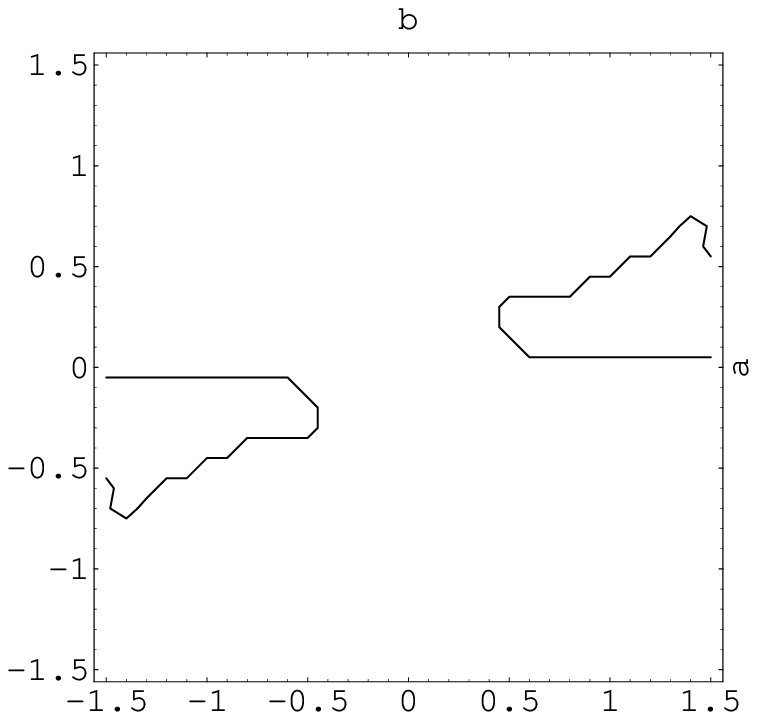,height=10cm,width=10cm,bbllx=2.cm,%
bblly=0.cm,bburx=13.cm,bbury=9.cm}}
\caption{\footnotesize Region (two disconnected parts)
in the $(b,a)$ parameter space for $M=8\times 10^9$ GeV where all mass    
splittings and mixing angles satisfy experimental constraints. (See text
for qualifications).
}
\end{figure}
%%%%%%%%%%%%%%%%%%%%%%%%figure%%%%%%%%%%%%%%%%%%%%%%%%

Figure 2, right plot, gives contour lines of constant $\Delta m_{23}^2$. 
The black (grey) region is excluded
because there $\Delta m_{23}^2<5\times 10^{-4}\, {\mathrm eV}^2$ ($\Delta
m_{23}^2>
10^{-2}\, {\mathrm eV}^2$), which is too small (large) to account for the
oscillations
of atmospheric neutrinos. Again, the white area is allowed. (The black area
corresponds in fact to the ``undecidable'' case discussed in the 
Introduction: it might be rescatable by unspecified extra effects.) 
We do not plot $\Delta m_{13}^2$ because it can always be inferred from
$\Delta m_{23}^2$ and  $\Delta m_{12}^2$. Moreover, in the interesting
case, $\Delta m_{12}^2\ll \Delta m_{23}^2$, one has $\Delta m_{13}^2
\simeq\Delta m_{23}^2$.

The intersection of the white areas in both plots is non-zero and would
give the allowed area concerning mass splittings. It is always the case
that the area surrounding the origin is excluded. There, the mass
differences are always of the same order, and follow the same pattern
discussed in subsection 2 ($\Delta m_{23}^2=2 \Delta m_{12}^2$). In any
case we conclude that, away from the origin, there is a non negligible
region of parameter space where it is natural to have $\Delta m_{23}^2\gg
\Delta m_{12}^2$ and in accordance with the values required to explain the
solar and atmospheric neutrino anomalies. In the following subsection we
explain the origin of this hierarchy of mass differences.

%%%%%%%%%%%%%%%%%%%%%%%%%figure%%%%%%%%%%%%%%%%%%%%%%%%
%\begin{figure}
%%%\psdraft
%\centerline{\hbox{
%\psfig{figure=cein6a.ps,height=10cm,width=10cm,bbllx=1.cm,%
%bblly=0.cm,bburx=12.cm,bbury=9.cm}
%\psfig{figure=cein6b.ps,height=10cm,width=10cm,bbllx=4.cm,% 
%bblly=0.cm,bburx=15.cm,bbury=9.cm}}}
%\centerline{\hbox{
%\psfig{figure=cein6c.ps,height=10cm,width=10cm,bbllx=1.cm,%
%bblly=0.cm,bburx=12.cm,bbury=9.cm}
%\psfig{figure=cein6d.ps,height=10cm,width=10cm,bbllx=4.cm,%
%bblly=0.cm,bburx=15.cm,bbury=9.cm}}}
%\caption
% \noindent{\footnotesize Same as figure 5 for
%different
%values of the Majorana mass: Upper: $M=10^{11}$ GeV; 
%Lower left: $10^{10}$ GeV; Lower right: $10^{9}$
%GeV.
%}
%\end{figure}
%%%%%%%%%%%%%%%%%%%%%%%%%figure%%%%%%%%%%%%%%%%%%%%%%%%

%%%%%%%%%%%%%%%%%%%%%%%%figure%%%%%%%%%%%%%%%%%%%%%%%%
\begin{figure}
%%\psdraft
\centerline{\hbox{
\psfig{figure=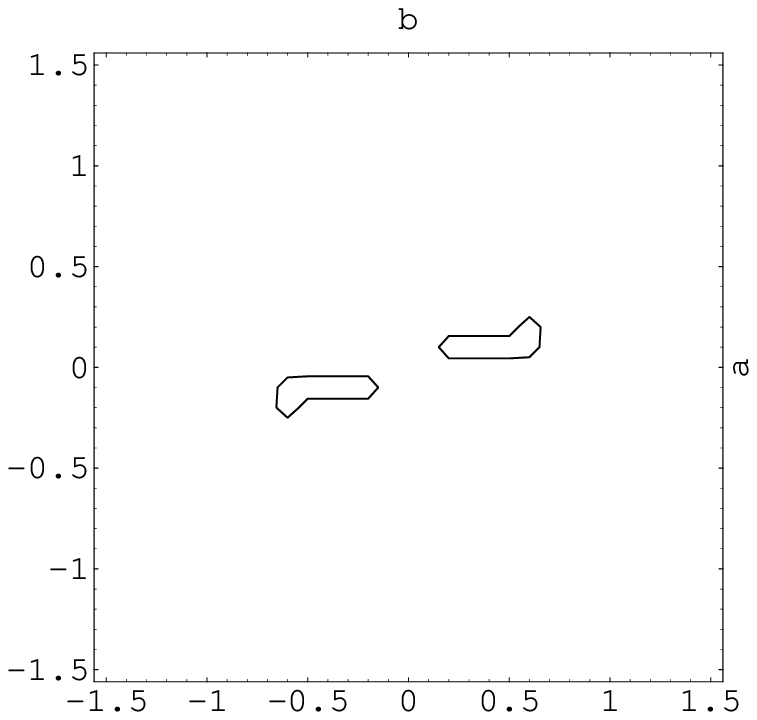,height=10cm,width=10cm,bbllx=2.5cm,% 
bblly=0.cm,bburx=13.5cm,bbury=9.cm}}}
\centerline{\hbox{
\psfig{figure=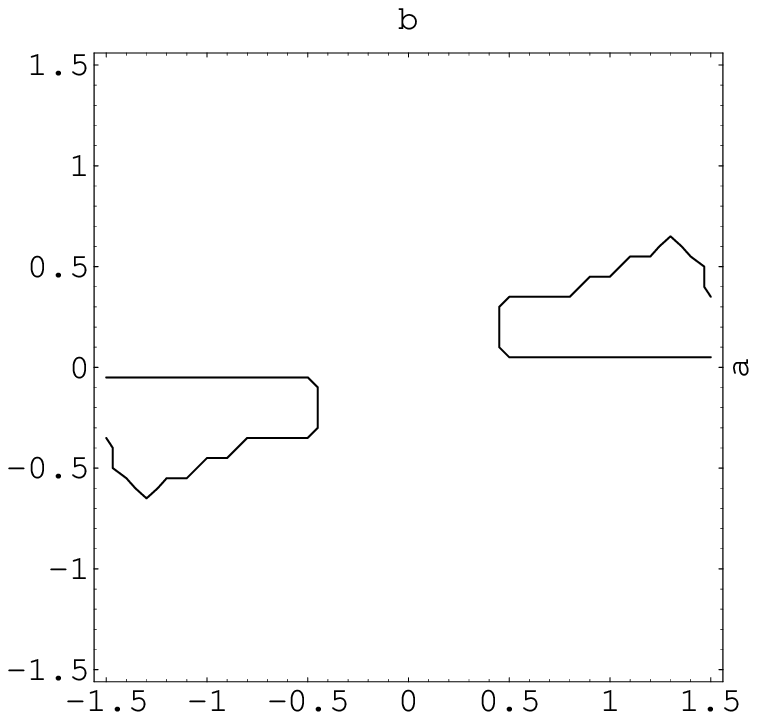,height=10cm,width=10cm,bbllx=1.cm,%
bblly=0.cm,bburx=12.cm,bbury=9.cm}
\psfig{figure=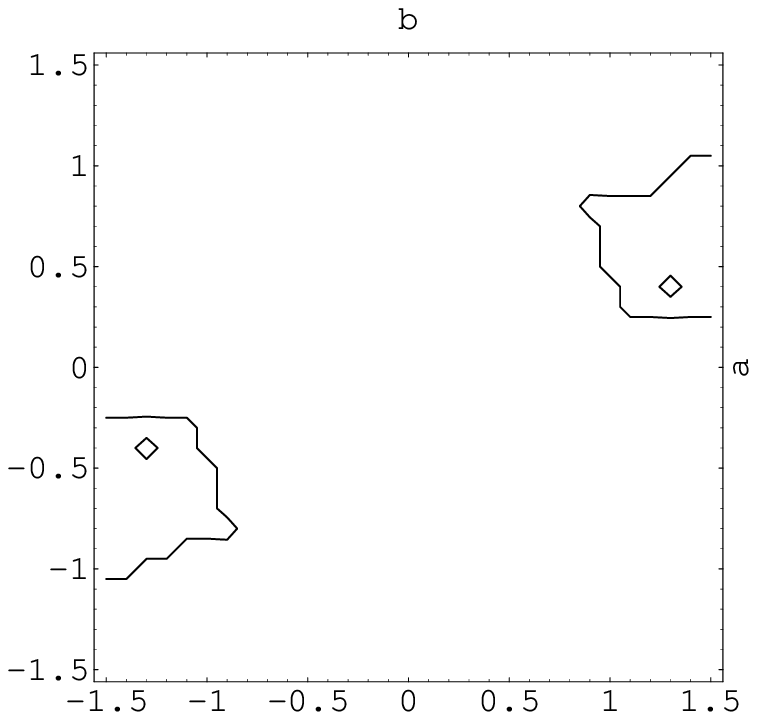,height=10cm,width=10cm,bbllx=4.cm,%
bblly=0.cm,bburx=15.cm,bbury=9.cm}}}
\caption
 \noindent{\footnotesize Same as figure 5 for
different
values of the Majorana mass: Upper: 
$10^{11}$ GeV; Lower left: $10^{10}$ GeV; Lower right: $10^{9}$
GeV.
}
\end{figure}
%%%%%%%%%%%%%%%%%%%%%%%%figure%%%%%%%%%%%%%%%%%%%%%%%%

Next, we need to ensure that the mixing angles are the correct ones to
give a good fit to atmospheric and solar neutrino data (as summarized by
the ranges given in the Introduction).  Figure 3, left plot, gives contours
of constant $\sin^22\theta_{2}$ (one of the mixing angles relevant for
atmospheric neutrino oscillations). The grey (white) area has
$\sin^22\theta_{2}$ larger (smaller) than 0.64 and is disfavored (favored)
by the data (SK $+$ CHOOZ) at 99\% C.L. according to the most recent analysis (last paper of 
ref.~\cite{range}).
%
%However, as explained in the Introduction, we do not 
%impose tight constraints on the value of
%this mixing angle. In any case, as we are about to see, the previous limit 
%does not constrain further the parameter space when the constraints on 
%the other mixing angles are considered.
%
The line singled-out corresponds to
$\sin^22\theta_{2}=0.36$ (maximum allowed value at 90\% C.L. according
to the same reference).

Figure 3, right plot, shows contours of constant $\sin^22\theta_{1}$ (the
other
mixing angle relevant for atmospheric neutrinos). The grey (white) area
corresponds to $\sin^22\theta_{1}$ smaller (larger) than 0.82, and is
thus disallowed (allowed). The additional line included has
$\sin^22\theta_{1}=0.9$.

Finally, figure 4, left plot, presents contours of constant 
$\sin^22\theta_{3}$ which
is relevant for oscillations of solar neutrinos. The grey (white) region
has $\sin^22\theta_{3}$ larger (smaller) than 0.99. If one is willing to
interpret the existing data as impliying an upper bound of 0.99 on
$\sin^22\theta_{3}$, then the grey region would be excluded. The plotted
curve gives $\sin^22\theta_{3}=0.95$ ($\sin^22\theta_{3}>0.8$ 
in all the region shown). Figure 4, right plot, shows the region of 
the parameter space accomplishing the resonance condition 
($\cos2\theta_{3}>0$), which is required for an efficient MSW solution
of the solar anomaly (see however the first paper of ref.~\cite{range} 
for caveats on this issue).

The region of parameter space where all constraints on mixing angles and
mass splittings are satisfied is given by the intersection of all white
areas in figures 2, 3 and 4 (right plot). 
If $\sin^22\theta_{3}<0.99$ is imposed, then
that intersection region, including now figure 4 (left plot), 
is empty and no allowed region remains. 
It should be noticed that this fact does not come from an incompatibility
between the previous constraint and the $\sin^22\theta_{3}>0.99$ obtained
from neutrinoless double $\beta$-decay limits, eq. (\ref{doublebeta}), in the
$\theta_2=0$ approximation. If this were
the case, it could be easily solved by decreasing the overall
size of the neutrino masses,  $m_\nu$, in eq.(\ref{doublebeta}), and this
is
not the case. Indeed,  eq.(\ref{doublebeta}) is satisfied in nearly all
the parameter space. Even where $\sin^22\theta_{3}<0.99$, this is still true
thanks to  the contribution of $\theta_2$.  What actually
forbids the whole parameter space
 if $\sin^22\theta_{3}<0.99$ is imposed,
is the incompatibility between acceptable $\theta_1$, $\theta_2$ and 
$\theta_3$ angles to fit simultaneously
all the neutrino oscillation data,
as can be seen from the figures. This fact remains when $m_\nu$ is
decreased.  In fact, the effect of decreasing $m_\nu$ is essentially an
amplification  of the figures shown here, which comes from the fact that
for a given Majorana mass, the neutrino Yukawa couplings become smaller
(the effect is similar to decrease $M$, which is discussed below).

If the $\sin^22\theta_{3}<0.99$
condition is relaxed (as discussed in the Introduction), 
then the allowed region is given by the two islands in figure
5, which is non-negligible.

If we now vary $M$, this allowed region will move in parameter space as
indicated in figure 6, where we show the allowed regions for
a sequence of Majorana masses that range from $10^9$ to
 $10^{11}$ GeV. As is apparent from the figure, lowering $M$ has the
effect of enlarging the allowed region which flies away from the origin,
leaving at some point the region of naturalness for $a$ and $b$. At
$M=10^8$ GeV there is no allowed region inside the natural range for
$(a,b)$. Conversely, increasing $M$ reduces the allowed region,
which gets closer to the origin (at $M=10^{12}$ GeV the allowed region
becomes extinct). 
 Let us remark again that in the allowed
region one fits the observed atmospheric and solar anomalies, while
part of the disallowed region corresponds in fact to the undecidable case
(more precisely the region near the $a=b=0$ origin),
in which some additional physics could be invocated to explain the same
data. In contrast, the region away from the origin becomes excluded
(the mass splittings are too large, even for atmospheric neutrinos).
Let us also notice that if the $\sin^22\theta_{3}<0.99$
 condition is imposed, the whole parameter space becomes disallowed 
for any value of $M$.

We also find that, whenever there is a hierarchy in the mass splittings,
the two lightest eigenvalues have opposite signs. This is just what is
needed to have a cancellation occurring in the neutrinoless double
$\beta$-decay constraint (\ref{doublebeta}). This constraint is satisfied in
almost the whole parameter space for any $M$.

This and other features of our results are
discussed further in the next subsection.

\subsection{Analytical understanding of results}

It is simple and very illuminating to derive analytical approximations
for the results presented in the previous subsection. The renormalization group
equations (\ref{rg1},\ref{rg2},\ref{rg3},\ref{rg4}) we integrated 
numerically can also be integrated
analytically in the approximation of constant right hand side. In this
approximation (which works very well for our analysis), the effective
neutrino mass matrix at low-energy is simply ${\cal M}_{b}$ plus some small
perturbation. It is straightforward to obtain how the degenerate eigenvalues of
${\cal M}_{b}$ get split by this perturbation. Neglecting the $Y_e$, $Y_\mu$
Yukawa couplings, we get the following analytical
expressions:
\bear{cl}
m_{\nu_1}\simeq &m_\nu \left[-1+
(2 c_a^2 c^2_b-1)\epsilon_\nu-2\epsilon_\tau
\right],\vspace{0.2cm}\\
m_{\nu_{2,3}}\simeq &m_\nu \left[1+3 \epsilon_\tau - c^2_a c^2_b
\epsilon_\nu\pm \left\{\left[\epsilon_\tau+(c^2_a c^2_b-c_{2a})\epsilon_\nu
\right]^2 +\left[s_{2a} s_b \epsilon_\nu - 2\sqrt{2}\epsilon_\tau 
\right]^2
\right\}^{1/2} \right]
\label{mnu123}
\eear
where $c_a=\cosh a$, $s_{2a}=\sinh 2a$, etc, and
\bea
\epsilon_\tau =\frac{Y_\tau^2}{128\pi^2}\left[\
\log\frac{M}{M_Z}+3\log\frac{M_p}{M}\right],
\eea
\bea
\epsilon_\nu=\frac{Y_\nu^2}{16\pi^2}\log\frac{M_p}{M}.
\eea
(The labelling of mass eigenvalues in eq.(\ref{mnu123}) may not always 
correspond to the conventional order $m_{\nu_1}^2<m_{\nu_2}^2<m_{\nu_3}^2$.)
It can be checked that, for $a=b=0$, the ${\bf Y}_{\nu}$ couplings play
 no role in
the mass splittings. This is not surprising since, as was mentioned in 
subsection
3.1, this case is equivalent to having all the structure in ${\cal M}$, while
${\bf Y}_{\nu}$ is proportional to the identity. Then, it can be seen from the
RGEs that all the non-universal modifications on ${\cal M}_{\nu}$ come from the
${\bf Y}_{e}$
matrix, and has a form similar to the one found in section 2 [see 
eq.(\ref{Mpert})]. So this scenario gives similar (not satisfactory) 
results to those found  in that section.

As $\epsilon_\nu\gg \epsilon_\tau$ (which occurs as soon as
${Y}_{\nu}>Y_\tau$, i.e 
for $M\simgt 10^9$ GeV), a further expansion in powers of
$\epsilon_\tau/\epsilon_\nu$ of the square root is possible in most of the
parameter space (except where the coefficient of $\epsilon_\nu^2$ inside that
square-root becomes very small). The mass eigenvalues then read
\bear{cl}
m_{\nu_1}\simeq &m_\nu \left[-1+
(2 c_a^2 c^2_b-1)\epsilon_\nu-2\epsilon_\tau
\right],\vspace{0.2cm}\\
m_{\nu_2}\simeq &m_\nu
\left[1-(2 c_a^2
c^2_b-1)\epsilon_\nu+\left(2-{\displaystyle\frac{1-c_{2a}-2\sqrt{2}
s_{2a}s_b}{c_a^2
c^2_b-1}}
\right)\epsilon_\tau\right]
,\vspace{0.2cm}\\
m_{\nu_3}\simeq &m_\nu
\left[1-\epsilon_\nu+\left(4+{\displaystyle\frac{1-c_{2a}-2\sqrt{2}
s_{2a}s_b}{c_a^2
c^2_b-1}}
\right)\epsilon_\tau\right].
\label{mnu1232}
\eear
These expressions show clearly the origin of the neutrino mass splittings. The
splitting between the first two neutrinos is controlled by the small parameter
 $\epsilon_\tau$, proportional to the squared Yukawa couplings of the charged
leptons,
and is insensitive to the parameter $\epsilon_\nu$ (proportional to the square of 
the larger Yukawa coupling $Y_\nu$)  which is responsible for the mass
difference of
the third neutrino.

From the previous expressions we can extract the following remarkable
conclusions.  If the neutrino Yukawa couplings are sizeable
(i.e. bigger than $Y_\tau$), we will automatically obtain a hierarchy
of mass splittings $\Delta m_{12}^2\ll \Delta m_{23}^2\sim \Delta
m_{13}^2$. This is exactly what is needed for a simultaneous solution
of the atmospheric and solar neutrino anomalies, and thus represents a
natural mechanism for the $\Delta m_{sol}^2\ll \Delta m_{at}^2$
hierarchy. Thus, if the mixing angle $\theta_2$ is small, as it turns
out to be in most of the parameter space, $\Delta m_{12}^2$ is to be
correctly identified with $\Delta m_{sol}^2$  and $\Delta m_{23}^2$ with
$\Delta m_{at}^2$. The two mass eigenvalues which are more degenerate
correspond to the lighter states, i.e.  $m_{\nu_1}^2\sim
m_{\nu_2}^2<m_{\nu_3}^2$. Moreover, $m_{\nu_1}$ and $m_{\nu_2}$ have
opposite signs in the diagonalized mass matrix, which is exactly what
is needed to fulfill the neutrinoless double $\beta$-decay condition,
eq.(\ref{doublebeta}).  All these nice features are illustrated by 
the explicit results  presented in the previous subsection.

Concerning the mixing angles, it is straightforward to check that,
working in the ${Y}_{\nu}>Y_\tau$ approximation, the eigenstates of the
perturbed ${\cal M}_\nu$ matrix, corresponding to the previous
eigenvalues, are
\bea
V_1' = V_1,\;\;\;V_2' = \frac{1}{\sqrt{\alpha^2+\beta^2}}
(\alpha V_2+\beta V_3),\;\;\;
V_3' = \frac{1}{\sqrt{\alpha^2+\beta^2}}(-\beta V_2+\alpha V_3),
\eea
where $V_i$ are the eigenstates corresponding to the bimaximal mixing
matrix,    
$V_b$
[see eq.(\ref{VGG})]
\bea
V_1 = \left (   
\begin{array}{c}
{\displaystyle \frac{-1}{\sqrt{2}}} \vspace{.1cm}\\
{\displaystyle\frac{1}{2}}  \vspace{.1cm}\\
{\displaystyle\frac{1}{2}} 
\end{array}  
\right ), ~~    
V_2 = \left (   
\begin{array}{c}
{\displaystyle{1 \over \sqrt{2}}} \vspace{.1cm}\\ {\displaystyle{1 \over
2}}  \vspace{.1cm}\\
{\displaystyle{1 \over 2}}
\end{array}  
\right ), ~~    
V_3 = \left (   
\begin{array}{c}
0  \vspace{.1cm}\\ {\displaystyle\frac{-1}{\sqrt{2}}}  \vspace{.1cm}\\
{\displaystyle\frac{1}{\sqrt{2}}}
\end{array}
\right ),
\label{vecbim}
\eea
and $\alpha, \beta$ are given by
\bea
\alpha=c_{a}s_b + O({\epsilon_\tau}/{\epsilon_\nu}),\;\;\;\;\beta=s_{a}
+ O({\epsilon_\tau}/{\epsilon_\nu}).
\label{alfabeta}
\eea
The $V_i'$ vectors define the new ``CKM'' matrix $V'$ from which the 
mixing angles are extracted. Clearly,  if   
just one of the two $(a,b)$ parameters is vanishing, then $V'=V_b$, i.e.
exactly the bimaximal mixing case.
Also, whenever $c_a$, $c_b$ are sizeable (i.e. away from $a=b=0$),
$|\alpha|\gg |\beta|$, and thus we are close to the bimaximal case.
Therefore, it is not surprising that in most of the parameter space shown
in the previous section, this was in fact the case. This is
remarkable, because it gives a natural origin for the bimaximal
mixing, which was not guaranteed a priori due to the ambiguity 
in the diagonalization of the
initial ${\cal M}_{\nu}(M_p)={\cal M}_{b}$ matrix, as was explained in
the Introduction.

On the other hand, the MSW condition
$\cos2\theta_{3}>0$ (written using the conventional 
order $m_{\nu_1}^2<m_{\nu_2}^2$) will be clearly satisfied as long as
$V_1$ corresponds to the lightest mass eigenvalue. In other words,
this condition requires that
the negative mass eigenvalue, see eqs.(\ref{mnu123},
\ref{mnu1232}),
corresponds to the lightest neutrino.

\subsection{Examples of acceptable ans\"atze}

At a generic point in the allowed regions we have found, the form of the
matrix ${\bf Y_\nu}(M_p)$  would look rather ad-hoc: different elements in
that matrix seem to conspire to give the correct neutrino mass texture.
However, there are particular cases in which this matrix has a plausible
structure. We give an example of such a texture for ${\bf Y_\nu}(M_p)$
which for the case $M\simeq 8\times 10^{9}$ GeV studied in previous
sections, would give mass splittings and mixing angles in agreement with
observation:
\bea
{\bf Y_\nu}(M_p) =  Y_\nu\,\pmatrix{
- {\displaystyle{1\over 2\sqrt2}} &  1  &  1 \cr
 {\displaystyle{1\over 2\sqrt2}}  &  1  &  1 \cr
       0  &  
- {\displaystyle{1\over\sqrt2}} &  {\displaystyle{1\over\sqrt2}} \cr
}.
\label{ansatz}
\eea
It corresponds to $a=0$ and $b=\sinh^{-1}(3/4)\simeq 0.69$, value which falls in
the allowed region plotted in figure 5. More precisely, the mass splittings are
\bea
\Delta m_{12}^2\simeq 2\times 10^{-5}\, {\mathrm eV}^2,\;\;\;
\Delta m_{13}^2\simeq 1\times 10^{-3}\, {\mathrm eV}^2,\;\;\;
\Delta m_{23}^2\simeq 1\times 10^{-3}\, {\mathrm eV}^2,
\eea
and the mixing angles
\bea
\sin^22\theta_{2}=0.04560,\;\;\;
\sin^22\theta_{1}=0.95405,\;\;\;
\sin^22\theta_{3}=0.99986.
\eea
Concerning the resonance condition for the MSW mechanism (see Fig.4), 
this ansatz lies precisely at the border of the allowed area.

Another examples of working ans\"atze can be obtained. 
For instance, the following ansatz (corresponding to $a=\cosh^{-1}(\sqrt{5}/2)
\simeq 0.48$, $b=\log(\sqrt{10}/2)\simeq 1.15$   )
\bea
{\bf Y_\nu}(M_p) =  Y_\nu\,\pmatrix{
- {\displaystyle {1\over 4}} &{\displaystyle{ \sqrt2}}   & 
{\displaystyle{3\over \sqrt2}}  \cr
 {\displaystyle{1\over 2\sqrt5}}  &  {\displaystyle{\sqrt5\over 2}}  &  
{\displaystyle{\sqrt5\over 2}} \cr
 - {\displaystyle{1\over 4\sqrt5}}    & 0 &  {\displaystyle{\sqrt5\over 2}} \cr
}.
\label{ansatz2}
\eea
works correctly for a wider range of right-handed Majorana masses
(e.g. for $M=10^{9}$ GeV and $M=10^{10}$ GeV, as can be seen from 
Figure 6).

It would be interesting to explore the possibility of finding a symmetry
that could be responsible for the form of these ans\"atze and to 
analyze their implications for future long baseline experiments
\cite{derujula}.

\section{Conclusions}

We have performed an exhaustive study of the possibility that
radiative corrections are responsible for the small mass 
splittings in the (cosmologically relevant) scenario of nearly degenerate
neutrinos. To do that, we assume that the initial   
form of the neutrino mass matrix (generated at high energy by
unspecified interactions) has the bimaximal mixing form, and
run down to low energy. We then examine the form of the low-energy
neutrino mass matrix,
checking its consistency with all the available experimental data,
including atmospheric and solar neutrino anomalies.

We find cases where the radiative corrections produce mass splittings
that are $i)$ just fine $ii)$ too large or $iii)$ too small.   
The vacuum oscillations solution to the solar neutrino problem
always falls in the situation $(ii)$, and it is therefore excluded.
On the contrary, if the initial bimaximal mass matrix is
produced by a see-saw mechanism (a possibility that we analyze
in a detailed way),
there are large regions of the parameter space
consistent with the large angle MSW solution, providing a natural origin
for the $\Delta m^2_{sol} \ll \Delta m^2_{atm}$ hierarchy.
Concerning the mixing angles, they are remarkably stable and close to the
bimaximal mixing form (something that is not guaranteed a priory,    
due to an ambiguity in the diagonalization of the initial matrix).
%For these nice results to appear, the rihgt-handed Majorana mass
%must lay in the range $10^8\ {\rm GeV} \leq M \leq 10^{12}\ {\rm GeV}$.

We have explained analytically the origin of these remarkable features,
giving explicit expressions for the mass splittings and the mixing angles.
In addition, we have presented particularly simple see-saw ans\"atze
consistent with all the observations.

Finally, we have noted that the scenario is very sensitive to a possible
upper bound on $\sin^2 2\theta_3$ (the angle responsible for the solar
neutrino oscillations). An upper bound
such as $\sin^2 2\theta_3<0.99$ would disallow completely the scenario
of nearly degenerate neutrinos due to the incompatibility between 
acceptable mixing angles to fit simultaneously all the neutrino 
oscillation data.

\section*{Addendum}

Shortly after completion of this work, a paper by J. Ellis and S. Lola
on the same subject appeared \cite{EL}. In it, the scenario of our
section~2 is also studied and similar (negative) conclusions reached. However,
as we show in our section 3, positive results are obtained when 
the general see-saw scenario as the mechanism responsible for the effective
neutrino mass matrix is studied.

Also, the treatment by these authors of the constraints on mixing angles
from neutrinoless double $\beta$-decay and LAMSW fits to solar neutrino
data is more pessimistic than ours.

\section*{Acknowledgements}

We thank D. Casper and E. Kearns for clarification of the results of
Super-Kamiokande fits.
This research was supported in part by the CICYT
(contract AEN95-0195) and the European Union
(contract CHRX-CT92-0004) (JAC). J.R.E. thanks the I.E.M. (CSIC, Spain)
and A.I, I.N the CERN Theory Division, for hospitality
during the final stages of this work. 

%%%%%%%%%%%%%%%%%%%%%%%%%%%%%%%%%%%%%%%%%%%%%%%%%%%%%%%%%%%%%%%%%%%

\end{document}